\documentclass[pdflatex,sn-mathphys-num,iicol]{sn-jnl}% Math and Physical Sciences Numbered Reference Style

\usepackage{graphicx}%
\usepackage{multirow}%
\usepackage{amsmath,amssymb,amsfonts}%
\usepackage{amsthm}%
\usepackage{mathrsfs}%
\usepackage[title]{appendix}%
\usepackage{xcolor}%
\usepackage{textcomp}%
\usepackage{manyfoot}%
\usepackage{booktabs}%
\usepackage{algorithm}%
\usepackage{algorithmicx}%
\usepackage{algpseudocode}%
\usepackage{listings}%
\usepackage[detect-none]{siunitx}

\newcommand{\et}{\ensuremath{E_\mathrm{T}}}
\newcommand{\etpred}{\ensuremath{E_\mathrm{T}^\mathrm{pred}}}
\newcommand{\ettrue}{\ensuremath{E_\mathrm{T}^\mathrm{true}}}
\newcommand{\etres}{\etpred-\ettrue}
\newcommand{\deltapred}{\ensuremath{\delta^\mathrm{pred}}}
\newcommand{\avmu}{\ensuremath{\langle\mu\rangle}}

\begin{document}

\title[Article Title]{Optimised neural networks for online processing of ATLAS calorimeter data on FPGAs}


\author*[1]{\fnm{Georges} \sur{Aad}}\email{aad@cern.ch}

\author[1]{\fnm{Rapha\"el} \sur{Bertrand}}\email{bertrand@cppm.in2p3.fr}

\author[1]{\fnm{Lauri} \sur{Laatu}}\email{lauri.laatu@cern.ch}

\author[1]{\fnm{Emmanuel} \sur{Monnier}}\email{monnier@in2p3.fr}

\author[2]{\fnm{Arno} \sur{Straessner}}\email{arno.straessner@tu-dresden.de}

\author[1]{\fnm{Nairit} \sur{Sur}}\email{nairit.sur@cern.ch}

\author[2]{\fnm{Johann C.} \sur{Voigt}}\email{johann\_christoph.voigt@tu-dresden.de}

\affil[1]{\orgdiv{CPPM}, \orgname{Aix-Marseille Universit{\'e}, CNRS/IN2P3}, \orgaddress{\city{Marseille}, \country{France}}}

\affil[2]{\orgdiv{Institut f{\"u}r Kern- und Teilchenphysik}, \orgname{Technische Universit{\"a}t Dresden}, \orgaddress{\city{Dresden}, \country{Germany}}}


\abstract{A study of neural network architectures for the reconstruction of the energy deposited in the cells of the ATLAS liquid-argon calorimeters under high pile-up conditions expected at the HL-LHC is presented. These networks are designed to run on the FPGA-based readout hardware of the calorimeters under strict size and latency constraints. Several architectures, including Dense, Recurrent (RNN), and Convolutional (CNN) neural networks, are optimised using a Bayesian procedure that balances energy resolution against network size. The optimised Dense, CNN, and combined Dense+RNN architectures achieve a transverse energy resolution of approximately \SI{80}{\mega\electronvolt}, outperforming both the optimal filtering (OF) method currently in use and RNNs of similar complexity. A detailed comparison across the full dynamic range shows that Dense, CNN, and Dense+RNN accurately reproduce the energy scale, while OF and RNNs underestimate the energy. Deep Evidential Regression is implemented within the Dense architecture to address the need for reliable per-event energy uncertainties. This approach provides predictive uncertainty estimates with minimal increase in network size. The predicted uncertainty is found to be consistent, on average, with the difference between the true deposited energy and the predicted energy.}




\maketitle
 
\section{Introduction}
\label{sec:introduction}

The ATLAS detector \cite{ATLAS:2008xda} at the Large Hadron Collider (LHC) \cite{Evans:2008zzb} measures the properties of particles produced in proton-proton (p-p) collisions that occur every \SI{25}{\nano\second} at each crossing of proton bunches. The collision energy can reach values of up to \SI{14}{\tera\electronvolt}. The liquid-argon (LAr) calorimeters are subsystems of the ATLAS detector. They are mainly designed to precisely measure the energy of electromagnetically interacting particles such as electrons and photons. Each of its \num{182468} cells produces an electronic pulse with an amplitude proportional to the energy deposited in the calorimeters. The pulses span a duration of around \SI{600}{\nano\second}. The LAr calorimeters produce several hundreds of terabits per second, necessitating specialized electronic boards to handle this enormous amount of data.

The LHC will be upgraded during a long shutdown period between 2026 and 2030 to reach the high-luminosity phase \cite{HL-LHC}. The High Luminosity LHC (HL-LHC) will produce up to 200 simultaneous p-p collisions in each bunch crossing (BC), which leads to signal pile-up. This imposes stringent requirements on the detectors to identify particles and measure their energy in this busy environment. Consequently, the LAr calorimeter electronics will be upgraded \cite{LAR-PHASE-II-TDR} during the same period (called the Phase-II upgrade) to improve its data processing capabilities. The new on-detector electronics will amplify, shape and digitise the detector pulses with a \SI{40}{\mega\hertz} sampling frequency. The digitised samples will be sent via optical fibers to off-detector electronic boards, called LAr Signal Processors (LASP), where they will be used to compute the deposited energy for each calorimeter cell. Each LASP board contains two processing units based on INTEL Agilex 7 FPGAs \cite{Agilex}. Each FPGA is expected to compute the energy of 384 calorimeter cells.

The LASP boards provide fast and reduced information to the ATLAS trigger system \cite{TDAQ-PHASE-II-TDR}, allowing for a rapid pre-selection of events flagged for further processing and permanent storage. This imposes stringent requirements on the latency allowed for the energy reconstruction algorithm. Currently, the latency of the energy reconstruction algorithm is required to be below \SI{125}{\nano\second}. Similar latencies are expected after the Phase-II upgrade. The LASP board also provides detailed information to the readout system for events selected by the trigger system. Currently, the optimal filtering (OF) algorithm \cite{OF} is used to compute the energy deposited in the calorimeters. The performance of the OF algorithm degrades significantly with the increased pile-up expected at the HL-LHC \cite{LAR-PHASE-II-TDR}, due to an increased number of overlapping pulses from energy deposits separated by less than 25 BCs.

The use of neural networks in FPGAs is a rapidly growing field, particularly in LHC experiments. The increasing processing power of FPGAs used during the Phase-II upgrade provides a unique opportunity to implement modern machine learning algorithms in the early stages of the data processing chain. Neural networks have been proposed to replace some of the trigger algorithms \cite{Duarte:2018ite,Bartz:2019dkp} after the Phase-II upgrade. Recently, machine learning algorithms have also been implemented in current trigger systems \cite{Govorkova:2021utb}. Neural networks are additionally being proposed for the first processing step to handle raw data from the detectors \cite{Sun:2022bxx}.

New neural network–based algorithms have been investigated to replace the OF algorithm for LAr data processing~\cite{LArNN1}. These algorithms have been shown to outperform the OF in terms of energy resolution in events with overlapping pulses due to pile-up, within an energy range between \SI{0}{} and \SI{5}{\giga\electronvolt}. For deployment in the LAr calorimeter readout, the neural networks must satisfy stringent requirements on latency and network size to match the computing resources available on the FPGAs, while still achieving a cell energy resolution superior to that of the OF. Neural networks of reduced size, implementing around 300 multiply–accumulate operations (MAC units), have already been successfully deployed on previous-generation INTEL FPGAs~\cite{LArNN2}. Given the increased resources and operating frequency of the Agilex 7 FPGA, a neural network with fewer than 500 MAC units is well within the capabilities of the LASP boards.

The study presented here extends previous work in several key aspects. It employs an improved dataset covering a larger dynamic range of the LAr calorimeters' analogue electronics compared to~\cite{LArNN1}, and uses a more realistic emulation of pulse overlap probability obtained by overlaying simulated inelastic proton–proton collisions. In addition, new neural network architectures are introduced that achieve optimal energy resolution under high pile-up conditions across the full dynamic range. The network hyperparameters are optimised using a Bayesian procedure that includes a penalty term on computational complexity, resulting in significantly smaller networks. Finally, this paper demonstrates how the per-event uncertainty on the energy predicted by the neural networks can be estimated with minimal additional computational cost by applying the Deep Evidential Regression technique~\cite{deep_evidential_regression}.

\section{Description of the network architectures}
\label{sec:nn_architectures}

Four neural network architectures are evaluated and compared with the OF algorithm. The networks predict the amplitude of the electronic pulse corresponding to an energy deposit in a calorimeter cell at a given BC, using digitised samples as input. The calorimeter pulse reaches its maximum between the second and third sample and subsequently decreases, developing a negative tail that extends over approximately 20 samples. The neural networks are trained with four post-deposit samples, starting from the BC of the targeted energy deposit, together with up to 28 pre-deposit samples to account for distortions induced by previous energy deposits. The OF algorithm, in contrast, uses five post-deposit samples and no pre-deposit samples.

The calorimeter response of a single cell is simulated with the AREUS \cite{AREUS-recent} toolkit. Isolated energy deposits from a hard-scattering event are overlaid on top of a continuous sequence of low-energy deposits representing pile-up arising from an average number of simultaneous p-p collisions per BC, $\avmu$, equal to 200. This corresponds to the worst-case scenario expected at the HL-LHC. For the simulation of the expected pulse sequence, a calorimeter cell in the middle layer of the barrel region at pseudorapidity $\eta=0.5125$ and azimuthal angle $\phi=0.0125$ is selected as a representative example \footnote{ATLAS uses a right-handed coordinate system with its origin at the nominal interaction point (IP) in the centre of the detector and the $z$-axis along the beam pipe. The $x$-axis points from the IP to the centre of the LHC ring, and the $y$-axis points upwards. Polar coordinates $(r,\phi)$ are used in the transverse plane,  $\phi$ being the azimuthal angle around the $z$-axis. The pseudorapidity is defined in terms of the polar angle $\theta$ as $\eta = -\ln \tan(\theta/2)$.}.

The neural networks are trained to target the true transverse energy \ettrue{} corresponding to the sum of energies deposited by both the hard-scatter and the in-time pile-up at a specific BC multiplied by $\frac{1}{\cosh\eta}$.
The simulated hard-scattering transverse energy is uniformly distributed between 0 and \SI{130}{\giga\electronvolt}, covering \SI{80}{\percent} of the digital dynamic range of the high-gain readout \cite{LAR-PHASE-II-TDR}. The low-gain setting, which extends the range to higher energies, is not considered here.

Unlike \cite{LArNN1}, this work concentrates on the energy reconstruction at specific BCs in which a hard-scatter has occurred. The results presented here assume an ideal trigger performance where hard-scatter events are perfectly detected\footnote{Cell energy reconstruction intended for trigger input requires the assignment of energy deposits to the correct p-p bunch-crossing by the neural network~\cite{LArNN1}, which is not considered here. Similarly, the OF algorithm would need to be enhanced by additional features such as finding peak maxima or including pulse timing information\cite{Aad:2022noz}.}. The focus is thus on the energy resolution and bias obtained in physics candidate events.

The training is performed with the Tensorflow Keras toolkit~\cite{tensorflow}. One million events are used for training, 1.5 million for validation, and 2.5 million for testing. The training sample size was found to be sufficient to reach optimal performance. Since the Bayesian optimisation (Section 3) uses the test sample to tune the hyperparameters, an independent sample of 13 million events is used for the final evaluation. The size of this latter sample was chosen to minimise statistical fluctuations across the transverse-energy ranges in which the networks are assessed. The detailed architectures and parameters of the networks are summarised in Table \ref{tab:nn_parameters}.

\begin{table*}[ht]
\centering
\caption{Summary of key parameters of the different neural networks. The number of samples corresponds to the sum of pre-deposit and post-deposit samples. For Dense and CNN architectures, “Layers” refers to the number of internal layers, while for RNNs it denotes the number of layers in the vanilla cell. The dimension is reported for each layer, with the kernel size additionally specified for the internal and output layers of the CNN. The parameters that are obtained from the Bayesian optimisation procedure described in Section~\ref{sec:bayesian_optimisation} are marked in boldface.}
\label{tab:nn_parameters}
\begin{tabular}{l|c|c|c|c}
\toprule
Architecture & RNN & Dense+RNN & Dense & CNN  \\
\hline
Input Samples & \textbf{1}+4 & \textbf{23}+4 & \textbf{28}+4 & \textbf{21}+4 \\
Internal Layers & 1 & 1 & 3 & 2 \\
Layers  Dimension & \textbf{8} & \textbf{5} & \textbf{10}$\,\oplus\,$\textbf{4}$\,\oplus\,$\textbf{8} & 19$\times$\textbf{5}$\,\oplus\,$9$\times$\textbf{6} \\
Kernel Size & - & - & - & \textbf{7}$\times$1$\,\oplus\,$\textbf{11}$\times$5$\,\oplus\,$\textbf{9}$\times$6 \\
Activation Function & ReLU & ReLU & ReLU & ReLU \\
\midrule
Number of Parameters & 89 & 161 & 392 & 431 \\
MAC units & 368 & 240 & 392 & 419 \\
\botrule
\end{tabular}
\end{table*}

\subsection{CNN architecture}

\begin{figure}
\centering
\includegraphics[width=1\linewidth]{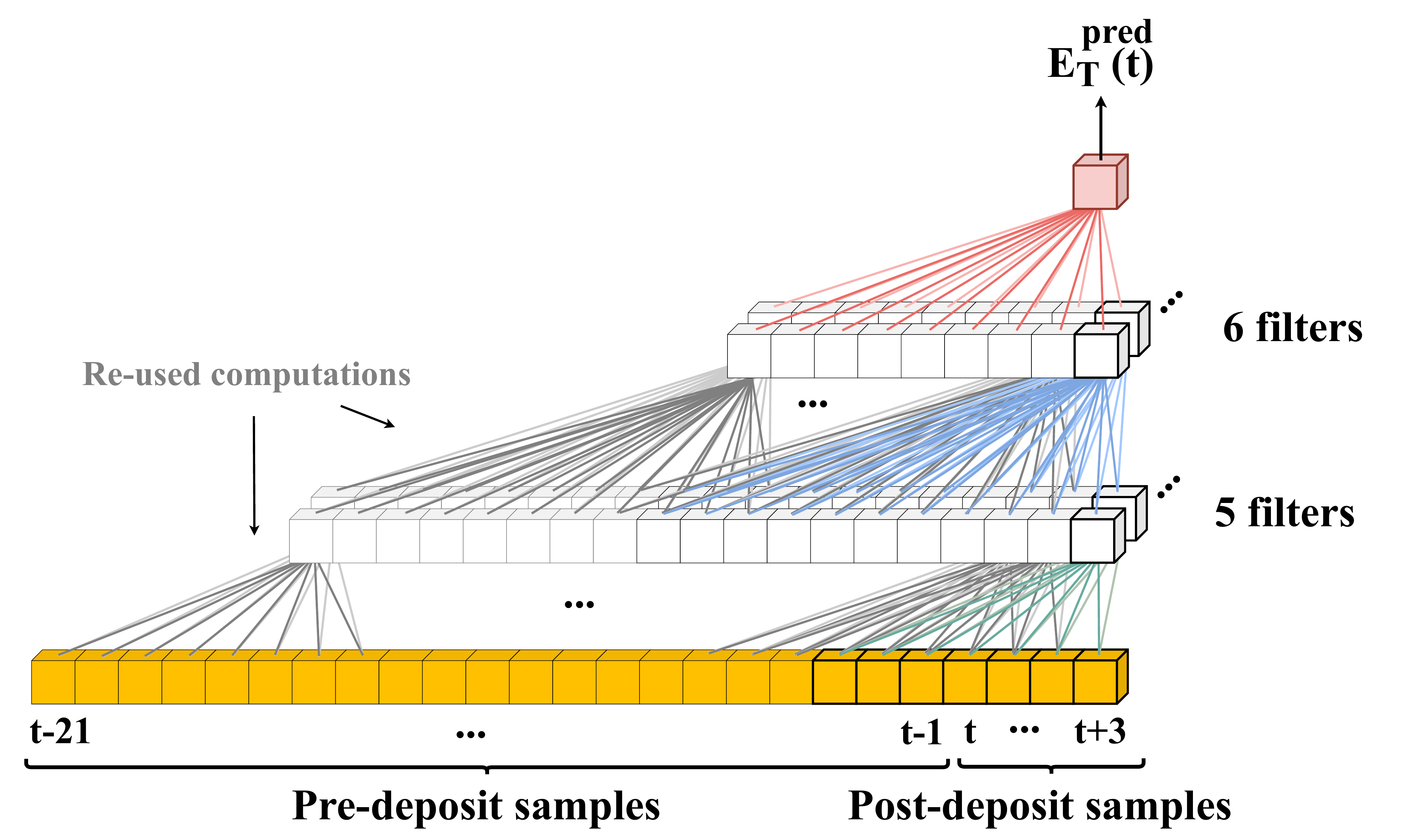}
\caption{The CNN architecture with 3 convolutional layers with kernels sliding in the time directions to compute the deposited energy at each BC. Only the final filter operations at each layer are computed at each BC, the previous filter results, represented with the grey connections, are reused from previous BCs}
\label{fig:cnn}
\end{figure}

The CNN architecture~\cite{cnnref} consists of two convolutional layers in addition to the input and output layers, as illustrated in Figure~\ref{fig:cnn}.
The parameters of this architecture, optimised through the Bayesian procedure described in Section~\ref{sec:bayesian_optimisation}, are listed in Table~\ref{tab:nn_parameters}.
The first convolutional layer applies five parallel one-dimensional filters with a kernel size of seven, sliding over 25 input samples in the time direction with a stride of one. The second convolutional layer applies six two-dimensional filters, each taking a 19$\times$5 input from the previous layer. These filters have a kernel size of 11$\times$5 and slide in the time direction with a stride of one. The output layer consists of a single filter with a kernel size of 9$\times$6, which is equivalent to a Dense layer with one neuron. All filters are followed by a rectified linear unit (ReLU) as activation function.

The CNN operates in a sliding-window mode over the input samples, computing the energy at each BC. In practice, only the final filter operation covering the most recent seven samples is explicitly computed for a given BC; earlier operations are identical to those already computed at the previous BC and can therefore be reused. This substantially reduces the number of MAC operations and lowers the algorithm's latency.

\subsection{RNN architecture}

\begin{figure}
\centering
\includegraphics[width=1\linewidth]{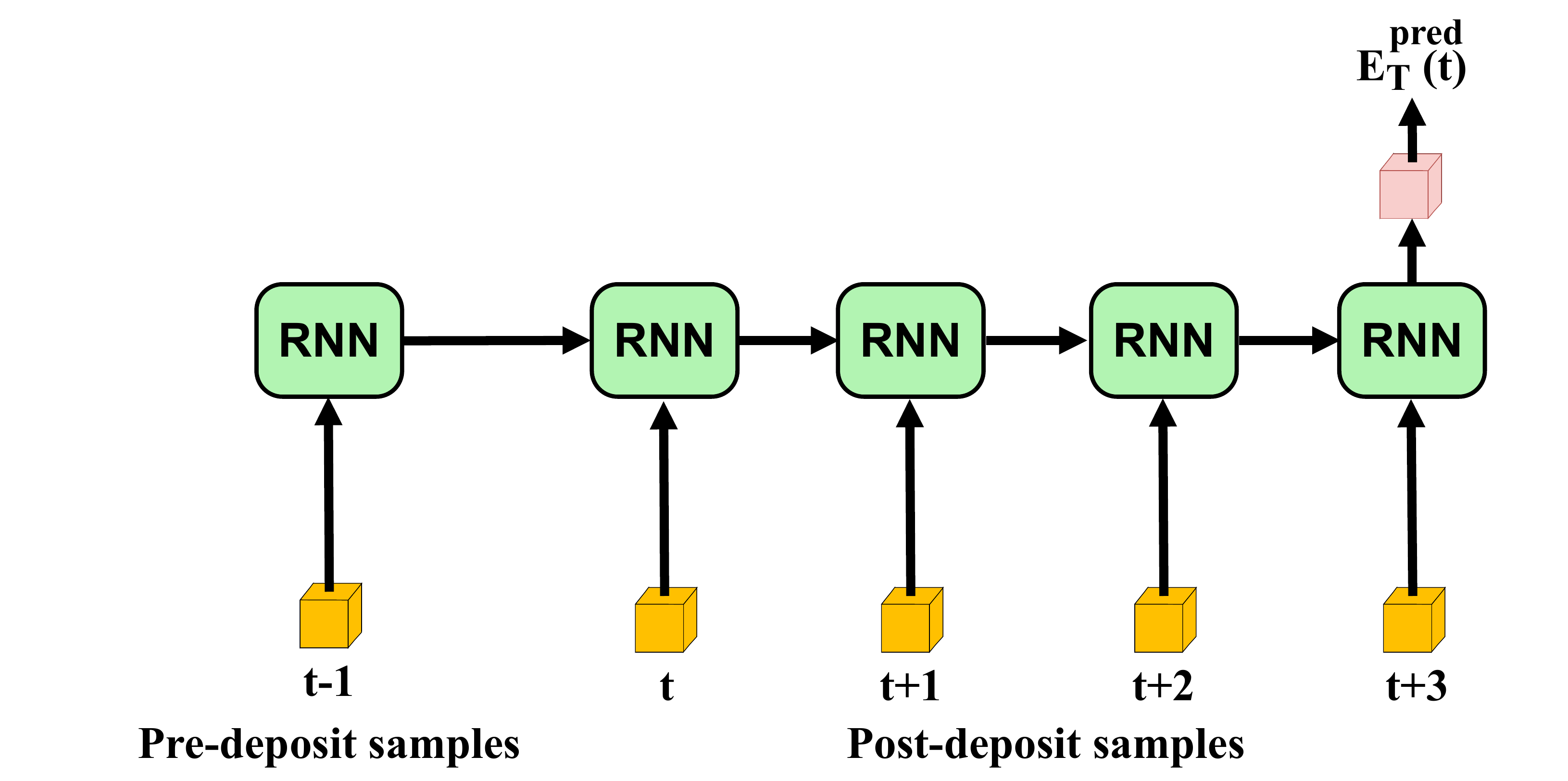}
\caption{The RNN architecture, where an RNN sequence is used to process the samples sequence, followed by a final dense layer computing the transverse energy}
\label{fig:rnn}
\end{figure}

The RNN~\cite{rnnref} architecture used in this work follows the design of Ref.~\cite{LArNN1}. RNNs are particularly well suited for time-series data, with computations performed synchronously with the arrival of data samples, thereby reducing latency. A sequence of RNN cells processes the digitised calorimeter pulse samples, each cell handling one sample at a given BC and combining it with the output of the previous cell as shown in Figure~\ref{fig:rnn}. To reduce network size, simple vanilla cells~\cite{vanillaref} are used, each consisting of a single layer of dimension eight followed by a ReLU activation. After the final sample arrives, only the last RNN cell and a dense layer computations are required to produce the result, minimizing latency. All five RNN cells share the same parameters; however, reuse of computations between BCs is limited because the first RNN cell is reset at each BC, leading to distinct inputs for each cell. In this study, the reuse of common operations is not considered when estimating the number of MAC units required for RNNs.

The network parameters are listed in Table~\ref{tab:nn_parameters}. The same values as in \cite{LArNN1} are used, since the optimisation described in Section~\ref{sec:bayesian_optimisation} did not yield improved results. Achieving competitive energy resolution requires a large RNN since its computational cost is proportional to the number of samples times the internal dimension squared while dense layers scale with the number of samples times the internal dimension. To address this, a modified RNN architecture is introduced in Section~\ref{sec:dense_rnn}.

\subsection{Dense+RNN architecture}
\label{sec:dense_rnn}

\begin{figure}
\centering
\includegraphics[width=1\linewidth]{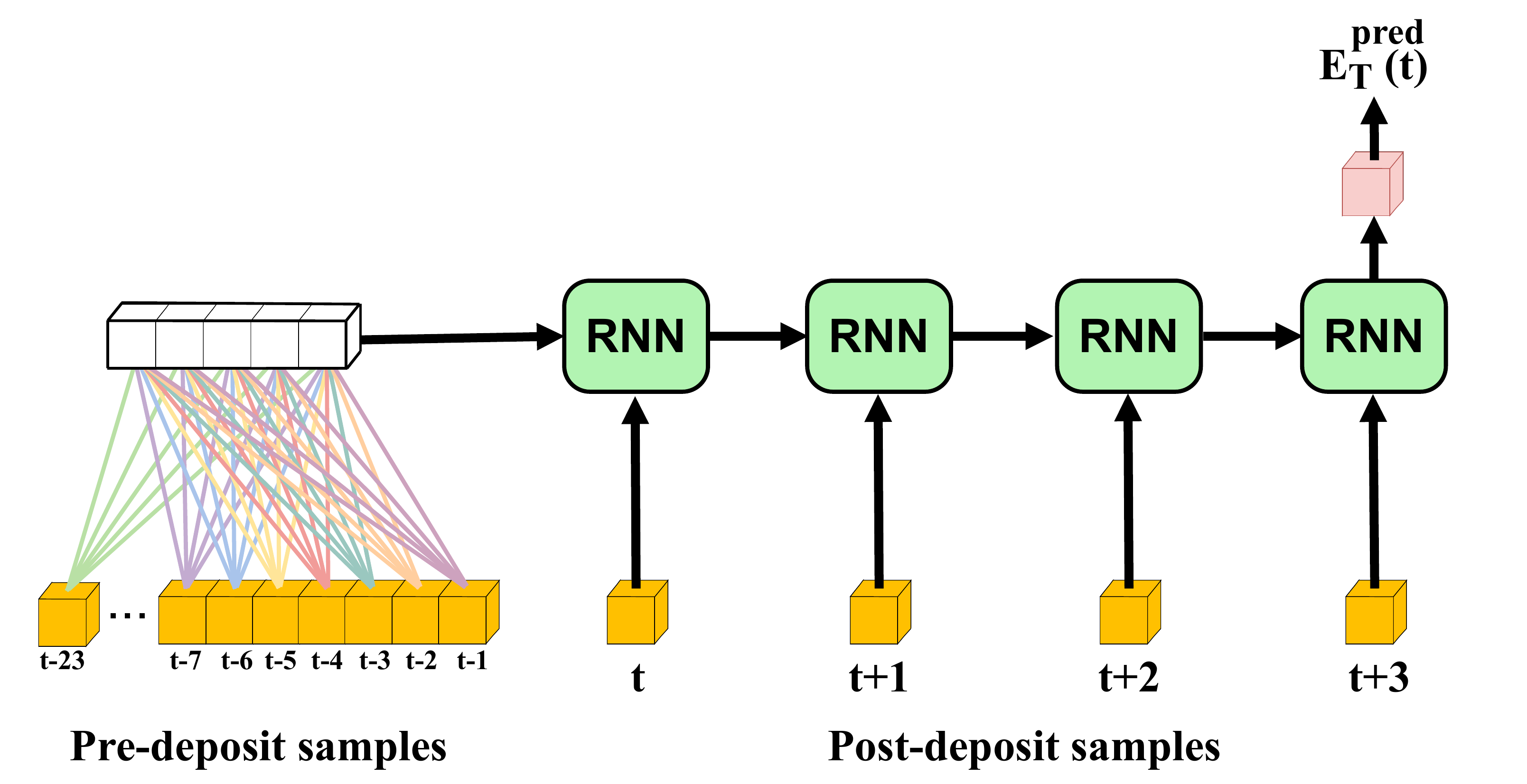}
\caption{The Dense+RNN architecture, where a dense layer processes pre-deposit samples to initialize an RNN sequence for post-deposit samples, followed by a final dense layer computing the transverse energy}
\label{fig:dense_rnn}
\end{figure}

The Dense+RNN architecture is designed to retain the advantages of RNNs while reducing computational cost for long sequences. Pre-deposit samples are processed by a dense layer, whose output initialises the RNN cell corresponding to the bunch crossing of the energy deposit, as illustrated in Figure~\ref{fig:dense_rnn}. A final dense neuron maps the last cell output to the transverse-energy measurement. The parameters of this architecture, optimised through the Bayesian procedure described in Section~\ref{sec:bayesian_optimisation}, are listed in Table~\ref{tab:nn_parameters}. Since the initial dense layer is applied only to pre-deposit samples and can be computed in advance, the design maintains a low latency.

\subsection{Staged Dense architecture}
\label{sec:dense}

\begin{figure}
\centering
\includegraphics[width=1\linewidth]{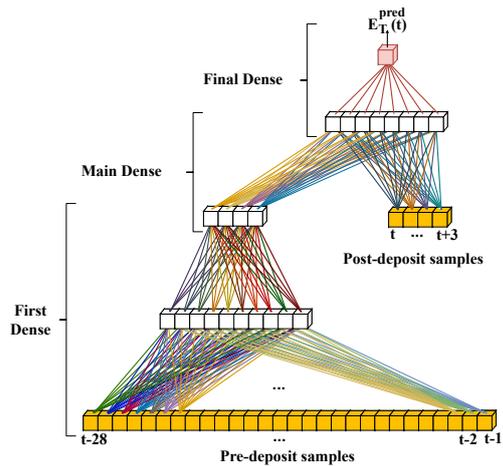}
\caption{The Staged Dense architecture, where a First  Dense block processes pre-deposit samples, the Main Dense combines this output with post-deposit samples, and the Final Dense outputs the transverse energy}
\label{fig:dense}
\end{figure}

The Staged Dense architecture (referred to as Dense in the following) further reduces the network size by using only dense layers. As shown in Figure~\ref{fig:dense}, input samples are incorporated in two stages to minimize latency. In the first stage, pre-deposit samples are used to correct for pulse distortions induced by previous energy deposits. These corrections are then combined, in a second stage, with post-deposit samples that capture the pulse shape from the targeted energy deposit. A final dense layer maps this information to the transverse-energy measurement. Latency is reduced since the first stage operates on pre-deposit samples and can be computed in advance. The parameters of this architecture, optimised through the Bayesian procedure described in Section~\ref{sec:bayesian_optimisation}, are listed in Table~\ref{tab:nn_parameters}.

\section{Bayesian optimisation of the network hyperparameters}
\label{sec:bayesian_optimisation}

The hyperparameters of the neural networks described in Section~\ref{sec:nn_architectures} are optimised to achieve the best possible energy resolution while keeping the network size minimal. The objective function, defined in Section~\ref{sec:objective_function}, balances energy resolution against network size. An iterative Bayesian optimisation procedure \cite{bayesian_optimization_first,bayesian_optimization_kernel} is used to determine the hyperparameter set that minimises this function. The optimised hyperparameters are the number of pre-deposit samples, the dimension of the internal layers for the Dense and RNN networks, and the kernel size and number of filters per layer for the CNN, while the number of layers and the number of post-deposit samples are kept fixed.

The optimisation proceeds in several steps: a surrogate distribution of the objective function is initialised with a flat prior, the objective function is then evaluated at one or several points in the hyperparameter space, and the surrogate is updated with the new observations. It begins with ten random evaluations and is repeated until the surrogate converges towards the objective function. In practice, 30 iterations are found to be sufficient to reach an optimal set of hyperparameters. In case of the CNNs, 100 iterations give more stable results, while the computational effort is still reasonable.

The surrogate is modelled with a Gaussian process using a 5/2 Matérn kernel, which has been shown to perform well for neural network optimisation \cite{bayesian_optimization_nn}. This approach provides both the mean and uncertainty of the surrogate at each point. The choice of evaluation points is critical for convergence. An acquisition function determines the next candidate point by balancing exploration (sampling high-uncertainty regions) and exploitation (sampling regions likely to yield optimal values) through a parameter $\xi$. The Expected Improvement criterion \cite{bayesian_optimization_ei} is employed to quantify the expected gain of each hyperparameter set over the current best hyperparameters.

To accelerate convergence, a multiprocess implementation evaluates several hyperparameter sets simultaneously at each iteration. A minimum Euclidean distance $d$ is enforced between sets to avoid redundant evaluations in the same local minimum. In addition, $\xi$ and $d$ are adjusted over three rounds of optimisation: the first round favours exploration with large $\xi$ and $d$ to locate distinct minima; the second round increases the density of sampled minima by gradually decreasing $d$ while keeping $\xi$ fixed; and the final round refines the hyperparameters around the identified minima by gradually reducing $\xi$ while fixing $d$ to its minimum value. This strategy significantly reduces the overall optimisation time. The corresponding framework is published in~\cite{code}.

\subsection{Hardware-constrained objective function}
\label{sec:objective_function}

The objective function used to optimise the neural network parameters balances two terms: the energy resolution and the number of MAC units, as shown in Equation~\ref{eq:objective_function}.

\begin{equation}
    f(M,\sigma) = 
    \begin{cases}
        \Tilde{\sigma} \text{\;~if\;} M \le 500\\
        \Tilde{\sigma}+0.3(\Tilde{M}-0.3) \text{\;~if\;} M \in \; ]500;850] \\
        \Tilde{\sigma}+0.3(\Tilde{M}-0.3)+e^{\Tilde{M}-0.65}-1 \text{\;else}
    \end{cases}
\label{eq:objective_function}
\end{equation}

Here, $M$ denotes the number of MAC units and $\sigma$ the energy resolution. $\Tilde{M}$ and $\Tilde{\sigma}$ represent the standardised forms of $M$ and $\sigma$, respectively, scaled to lie between zero and one. The standardisation procedure ensures that the two terms have comparable magnitudes and is defined in Equation~\ref{eq:standardization}.

\begin{equation}
    \Tilde{\sigma} = \frac{\sigma-\SI{0.07}{\giga\electronvolt}}{\SI{1.13}{\giga\electronvolt}}, \quad
    \Tilde{M} = \frac{M-200}{1000}
\label{eq:standardization}
\end{equation}

No penalty on the network size is applied below 500 MAC units, a linear penalty is introduced above 500, and an additional exponential penalty beyond 850, where FPGA implementation becomes increasingly difficult. The constant parameters in Equation~\ref{eq:objective_function} are chosen to ensure that the function is continuous and are tuned heuristically to achieve the best performance. Since neural network training is not deterministic and depends on random weight initialisation, five trainings are performed for each parameter set, and the configuration with the best objective score is selected.

\subsection{Results}
\label{sec:bayesian_results}

\begin{figure}
\centering
\includegraphics[width=1\linewidth]{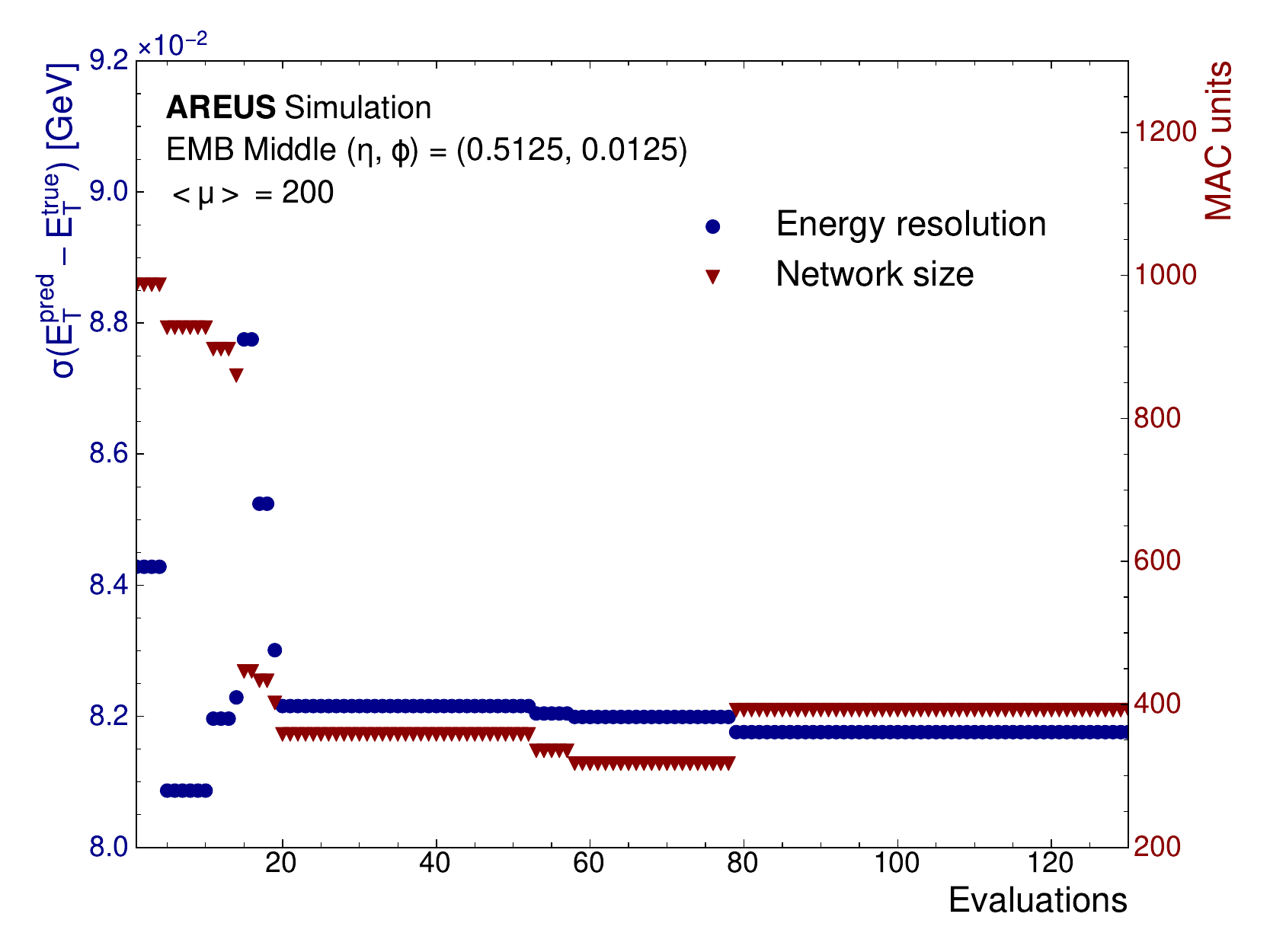}
\caption{Transverse energy resolution and number of multiply–accumulate (MAC) units for the Dense neural network architecture as a function of the evaluation number corresponding to a given hyperparameter set in an iterative Bayesian optimisation process. The best model up to each evaluation is shown, selected based on a score balancing energy resolution with the number of MAC units to achieve an optimal trade-off between resolution and network size. The resolution is computed for a single cell of the LAr calorimeter barrel section (EMB) for a dataset with pile-up of $\avmu=200$}
\label{fig:baysesian_optimisation}
\end{figure}

The evolution of the energy resolution and the number of MAC units during the Bayesian optimisation procedure for the Dense architecture is shown in Figure~\ref{fig:baysesian_optimisation}. For each of the 30 iterations, four hyperparameter sets are evaluated simultaneously, leading to 120 evaluations in addition to 10 random sets at the beginning of the procedure. Figure~\ref{fig:baysesian_optimisation} displays the best model reached up to a given evaluation, selected according to the objective function score. After the ten random sets, the best network size is drastically reduced, while the energy resolution initially degrades. The resolution is recovered after about 20 evaluations without significant change in the network size. In the subsequent 100 evaluations, only small improvements in the best network size and resolution are observed.

Because both the optimisation and the training are stochastic, several networks with similar resolutions and MAC counts emerge but differ in performance across the energy range. Among these, the network with the flattest energy-scale and resolution distributions over the full \ettrue{} range is selected.

The optimised hyperparameters and the corresponding number of MAC units are listed in Table~\ref{tab:nn_parameters}. All networks use fewer than 500 MAC units, making their implementation on the FPGAs of the LAr system feasible. All optimised networks, except the RNN, employ more than 20 pre-deposit samples, enabling them to efficiently capture distortions from previous energy deposits. RNNs with long sequences, however, become too large to fit into FPGAs. The Bayesian optimisation procedure did not produce a better performing RNN than the one originally used in \cite{LArNN1}. Therefore, the same RNN as in \cite{LArNN1} is adopted here.

\begin{figure}
\centering
\includegraphics[width=1\linewidth]{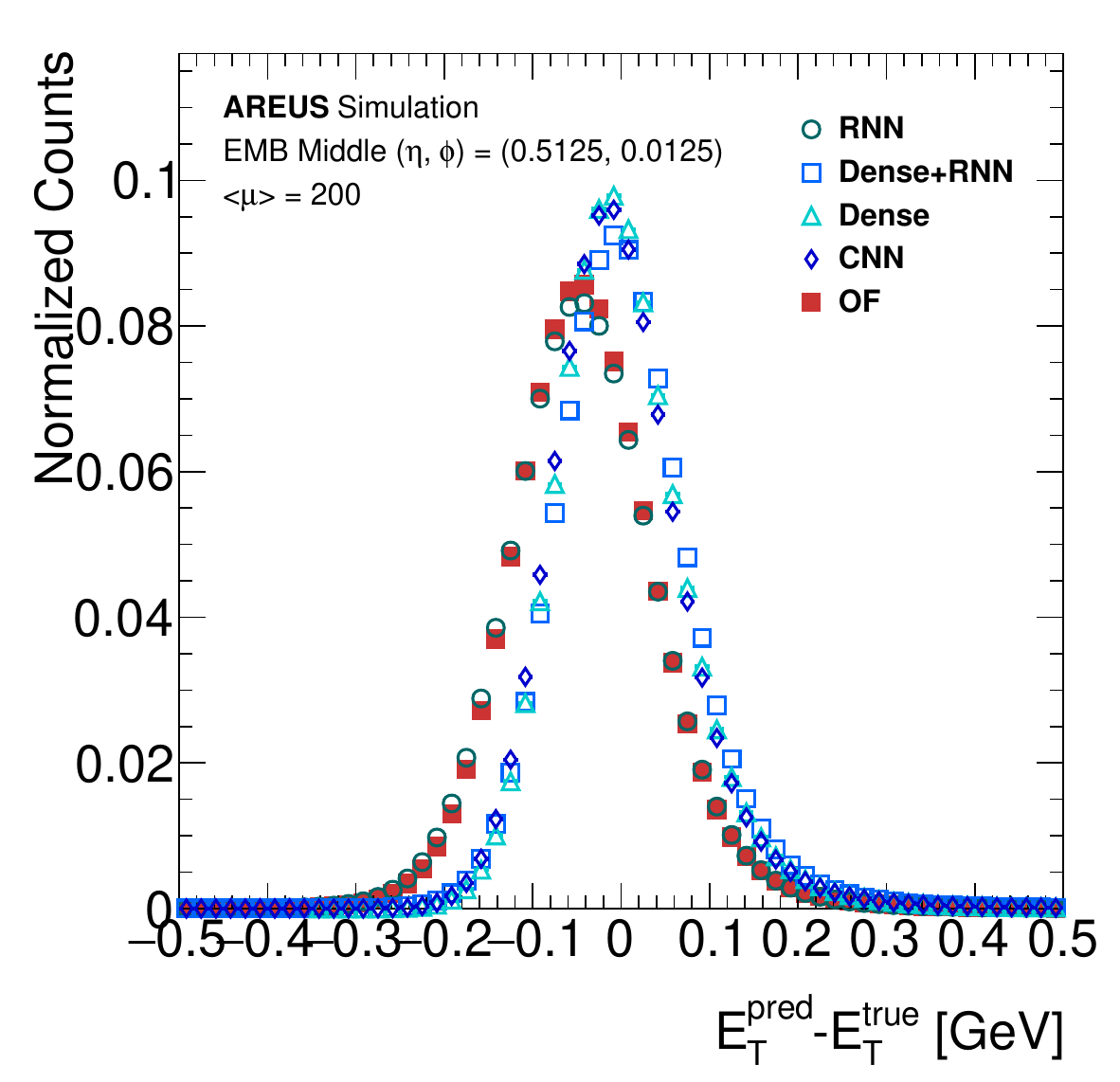}
\caption{Difference between the predicted and true transverse energy deposited in a single cell of the liquid-argon barrel calorimeter (EMB) for a dataset with a pile-up of $\avmu=200$. The four neural network algorithms are compared with the legacy optimal filtering algorithm}
\label{fig:et_resolution}
\end{figure}

The transverse energy resolution is evaluated in terms of the difference between the predicted and true transverse energy (\etres) obtained for the neural networks, and compared to the OF, as shown in Figure~\ref{fig:et_resolution}. The Dense+RNN, Dense, and CNN architectures achieve similar resolutions, all outperforming the RNN and OF. It can also be seen that the OF and the RNN systematically underestimate the transverse energy, whereas the other networks reconstruct values consistent with \ettrue. This behaviour is expected for the OF, which is designed to estimate only the injected energy without the average in-time pile-up component. The RNN, with its reduced number of input samples, cannot capture long-range dependencies in the pulse and thus fails to outperform the OF.

\begin{figure}
\centering
\includegraphics[width=1\linewidth]{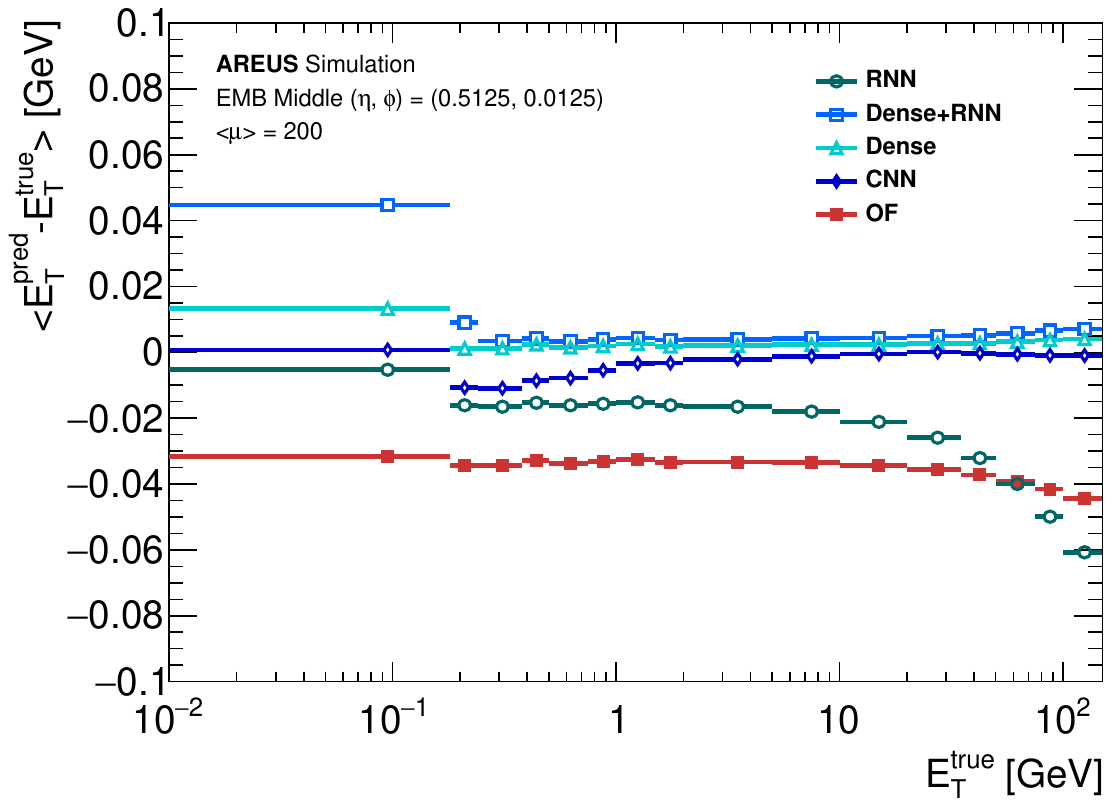}
\caption{Transverse energy scale, represented by the mean of the difference between the predicted and true transverse energy, as a function of the true transverse energy deposited in a single cell of the liquid-argon barrel calorimeter (EMB) for a dataset with a pile-up of $\avmu=200$. The four neural network algorithms are compared with the legacy optimal filtering algorithm}
\label{fig:et_resolution_mean}
\end{figure}

To further study performance over the full dynamic range, \etpred{} is compared with \ettrue{} in bins of \ettrue{}. The transverse energy scale is probed using the mean of \etres{} as a function of \ettrue{} in Figure~\ref{fig:et_resolution_mean}. As expected, the OF underestimates the energy across the full range. The Dense and Dense+RNN slightly overestimate the energy, while the CNN slightly underestimates it, particularly at low \ettrue{}. The flat behaviour of the mean of \etres{} for the Dense+RNN, Dense, and CNN demonstrates that these networks capture the energy scale across the full dynamic range. The RNN slightly underestimates the energy at low \ettrue{}, with the discrepancy growing significantly at high \ettrue{}, where the OF outperforms the RNN.

\begin{figure}
\centering
\includegraphics[width=1\linewidth]{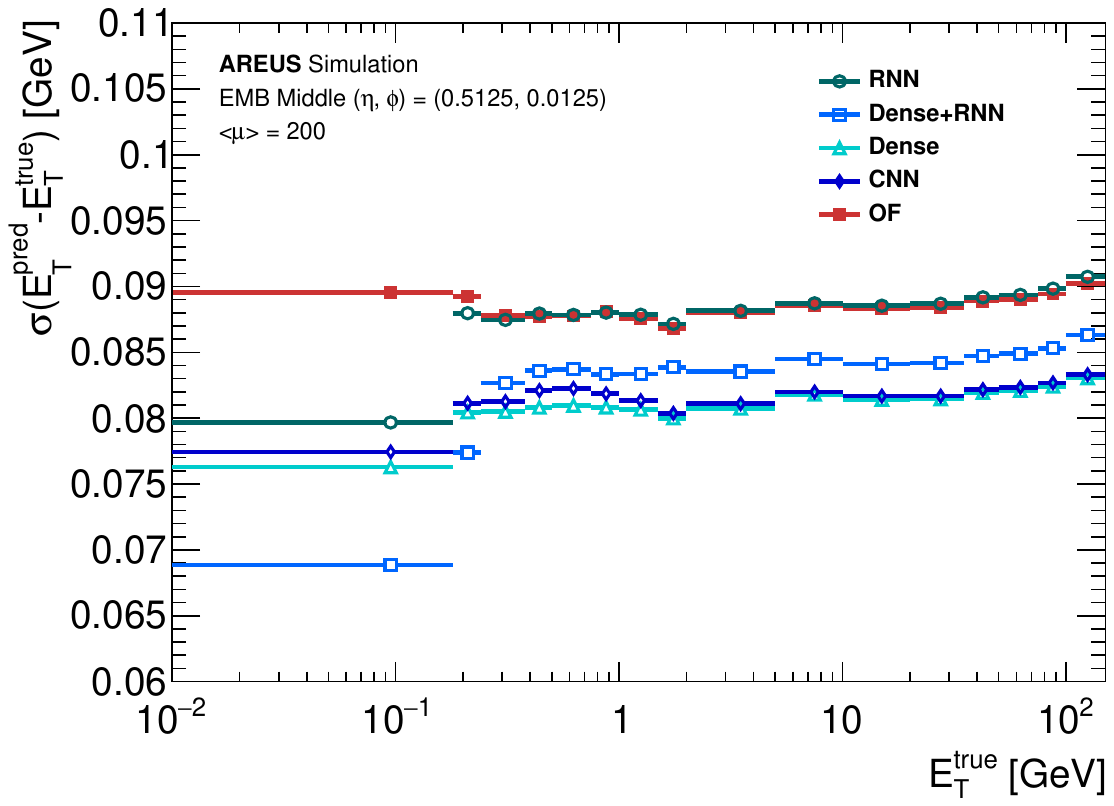}
\caption{Transverse energy resolution, represented by the standard deviation of the difference between the predicted and true transverse energy, as a function of the true transverse energy deposited in a single cell of the liquid-argon barrel calorimeter (EMB) for a dataset with a pile-up of $\avmu=200$. The four neural network algorithms are compared with the legacy optimal filtering algorithm}
\label{fig:et_resolution_std}
\end{figure}

The transverse energy resolution is quantified using the standard deviation of \etres{} as a function of \ettrue{} in Figure~\ref{fig:et_resolution_std}. The resolution is nearly flat across the full \ettrue{} range for all algorithms. The Dense and CNN architectures show the best performance, with an energy resolution of around \SI{80}{\mega\electronvolt}. The RNN and OF achieve around \SI{90}{\mega\electronvolt}.

\newpage

\section{Event-wise computation of the energy uncertainty}

The uncertainty in the computed energy deposited in a LAr calorimeter cell is of great importance for the subsequent steps of data acquisition and reconstruction. The digitised samples are transmitted to readout only for cells with significant energy above a noise threshold, which corresponds to a precomputed average energy uncertainty in a given cell. In addition, only cells with significant energy are used as seeds for the clustering algorithms \cite{ATLAS:2016krp} employed by ATLAS to build calorimeter clusters and, subsequently, physics objects. The uncertainty in the cell energy also propagates to the uncertainty in the reconstructed energy of physics objects, notably electrons and photons \cite{ATLAS:2018krz}. Currently, the uncertainty is determined as the sum of the electronic noise and the average expected pile-up noise for a given cell. These uncertainties are updated only when a significant change in \avmu{} is expected. They do not account for instantaneous luminosity variations during an LHC run, differences in luminosity between BCs due to the LHC bunch-filling structure, or event-by-event fluctuations reflecting the actual preceding deposits in a calorimeter cell. A per-event computation of the uncertainty could allow all these effects to be taken into account, providing a more accurate estimate, especially in high-pile-up environments.

Standard regression networks output only a single deterministic value, offering no information about the confidence or reliability of the prediction. Bayesian neural networks (BNNs) are capable of estimating uncertainties but are computationally expensive: they treat weights as random variables and require repeated posterior sampling, making their implementation on the FPGAs of the LAr Phase-II hardware infeasible. Deep Evidential Regression (DER), on the other hand, offers a practical approach for learning uncertainty without requiring sampling during training or inference. It does so by constructing a probabilistic distribution over the network output using a Normal–Inverse-Gamma (NIG) distribution, which enables the inference of both the predicted value and its associated uncertainties. The uncertainty is decomposed into aleatoric uncertainty, arising from intrinsic noise in the data, and epistemic uncertainty, arising from limited knowledge or model capacity.

The NIG distribution is parameterised by four values: $\gamma$, representing the expectation value of the predicted transverse energy ($\mathbb{E}[\etpred]$); $\nu$, controlling the confidence in this prediction (epistemic variance, $\text{Var}[\etpred]$); and $\alpha$ and $\beta$, which shape the distribution of the variance (aleatoric variance, $\mathbb{E}[\sigma^2]$), as outlined in Equation~\ref{eq:uncertainty}. Here, $\delta^{pred}$ is the total predictive uncertainty, computed as the sum in quadrature of the aleatoric and epistemic contributions.

\begin{equation}
\begin{aligned}
    \mathbb{E}[\etpred] &= \gamma \\
    \text{Var}[\etpred] &= \frac{\beta}{\nu (\alpha - 1)} \\
    \mathbb{E}[\sigma^2] &= \frac{\beta}{\alpha - 1} \\
    \delta^\mathrm{pred} &= \sqrt{\text{$\mathbb{E}$}[\sigma^2]+\text{$\textrm{Var}$}[\etpred]}
\end{aligned}
\label{eq:uncertainty}
\end{equation}

\subsection{Network architectures with deep evidential regression}

The staged dense architecture described in Section~\ref{sec:dense} is modified to incorporate DER. The final dense layer, originally consisting of a single neuron, is substituted by a \textit{DenseNormalGamma} layer from the evidential deep learning Python library \cite{der_github}. This layer is formally equivalent to a standard dense layer with four output neurons corresponding to the parameters of the NIG distribution: $\gamma$, $\nu$, $\alpha$, $\beta$. All other parameters of the dense network are kept unchanged. The loss function for the DER network is defined as the sum of a negative log-likelihood term, which constrains the data to the evidential distribution, and a regularisation term, which penalises the assignment of high evidence to incorrect predictions, as detailed in \cite{deep_evidential_regression}.

\subsection{Results}

The energy resolution and associated uncertainties are computed using the Dense network with the DER technique. The resulting resolution is compared with that obtained using the Dense architecture described in Section~\ref{sec:bayesian_results}. Integrated over the full \ettrue{} range, the resolution is \SI{82}{\mega\electronvolt} both with and without DER. No significant degradation is observed across the full range.

\begin{figure}
\centering
\includegraphics[width=1\linewidth]{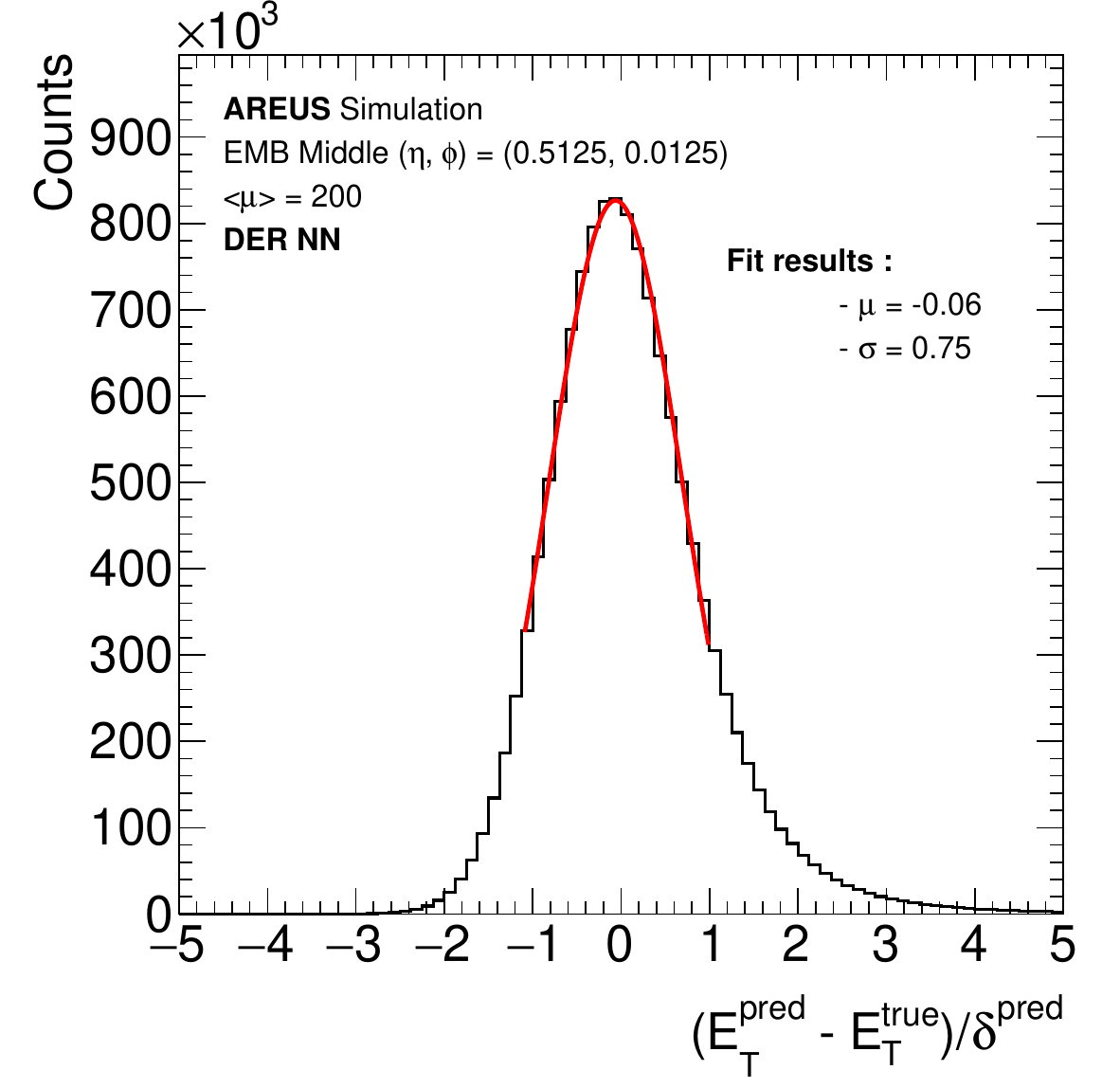}
\caption{Pull distribution illustrating the distribution of the difference between the predicted and true transverse energy deposited in a single cell of the liquid-argon barrel calorimeter (EMB), divided by the predicted uncertainty ($\delta^{\textrm{pred}}$), for a dataset with a pile-up of $\avmu=200$. The predicted transverse energy and its associated uncertainty are computed using a Neural Network with a dense architecture, employing the Deep Evidential Regression technique}
\label{fig:pull}
\end{figure}

Figure~\ref{fig:pull} shows the pull distribution, defined as (\etres)/\deltapred. A Gaussian fit to the core yields a mean close to zero and a standard deviation of 0.75, indicating a slight overestimation of the predicted uncertainty. The distribution also exhibits a slight asymmetry towards positive values, corresponding to a tendency for \etpred{} to be overestimated. Overall, the network provides uncertainty estimates that are, on average, consistent with the deviations between predicted and true \et.

\begin{figure}
\centering
\includegraphics[width=1\linewidth]{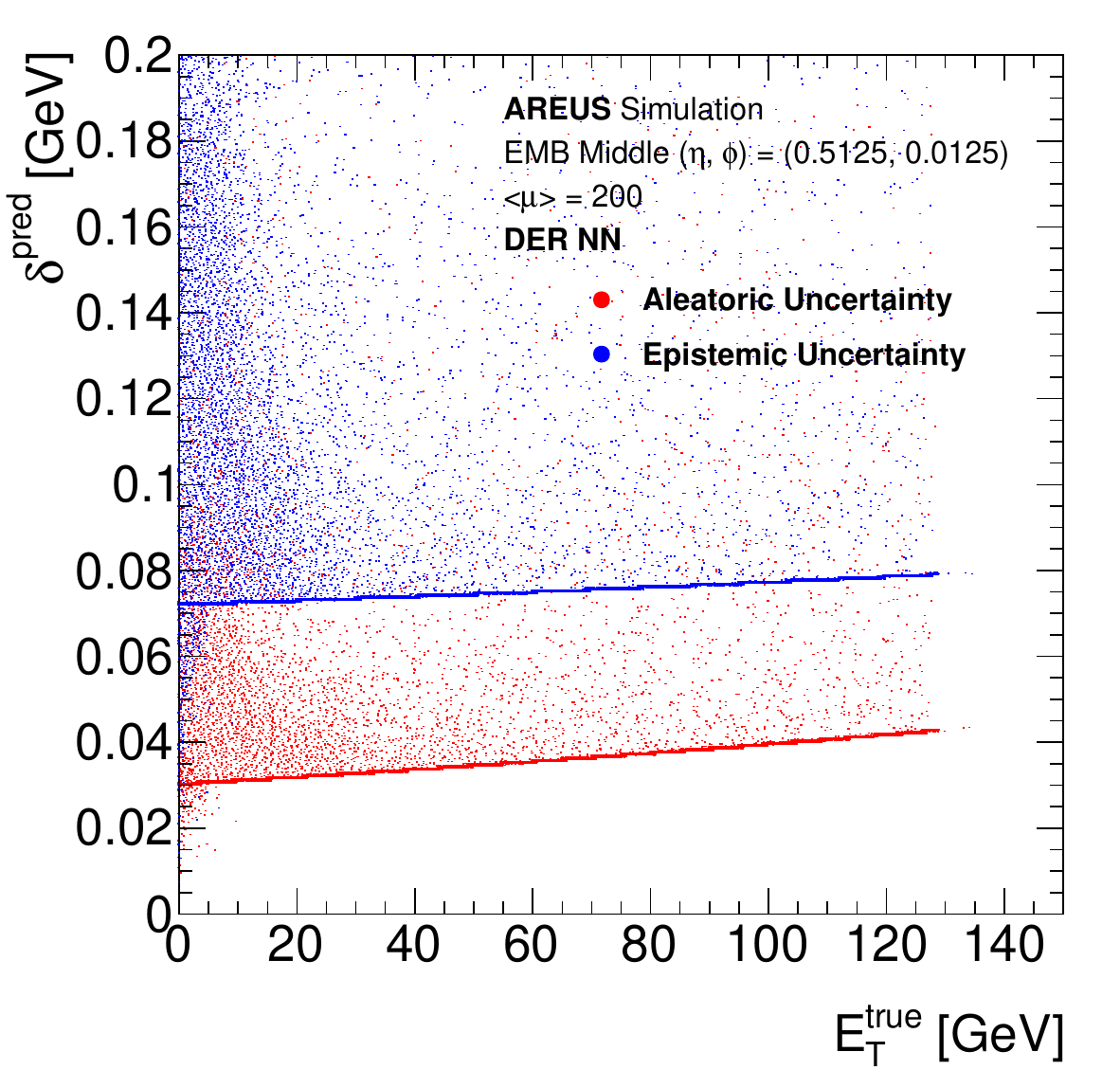}
\caption{Aleatoric and epistemic uncertainties as functions of the true transverse energy deposited in a single cell of the liquid-argon barrel calorimeter (EMB), for a dataset with a pile-up of $\avmu=200$. The uncertainties are obtained with the Dense neural network architecture employing the Deep Evidential Regression technique}
\label{fig:der_uncertainties}
\end{figure}

Figure~\ref{fig:der_uncertainties} shows the aleatoric and epistemic uncertainties as functions of \ettrue{}. The epistemic component is dominant, suggesting that further improvements could be achieved through additional optimisation of the network architecture or an increase in the training dataset size. The uncertainty distributions are sharply peaked around a single value for both components. Approximately 99\% of events fall within a narrow epistemic uncertainty band ranging from \SI{72}{\mega\electronvolt} at low \ettrue{} to \SI{79}{\mega\electronvolt} at high \ettrue{}. Similarly, 95\% of events lie within a narrow aleatoric uncertainty band ranging from \SI{30}{\mega\electronvolt} at low \ettrue{} to \SI{42}{\mega\electronvolt} at high \ettrue{}. This behaviour is expected, since the energy resolution distribution is approximately Gaussian, with very small tails corresponding to significant deviations of \etpred{} from the true deposited energy. About 1\% of events exhibit \etres{} values exceeding three standard deviations of the energy resolution. This fraction is too small for the network to reliably identify such events as outliers and assign them large \deltapred{} values. As a result, the correlation between \etres{} and \deltapred{} remains weak. Consequently, \deltapred{} cannot be directly used to reject poorly reconstructed events. However, these events are negligible in this sample, and \deltapred{} accurately captures the core energy resolution.

Future work will explore the application of this method in scenarios where \deltapred{} can be used to reject large-resolution tails caused by noise bursts in the detector or by a non-uniform bunch structure in the LHC.

\section{Conclusion}

The upgrade of the liquid-argon (LAr) readout electronic boards provides a unique opportunity to deploy modern machine learning algorithms, embedded on FPGAs, to improve the reconstruction of the energy deposited in the LAr calorimeters. Several neural network architectures were tested on a simulated sample of energy deposits covering nearly the full dynamic range of a single cell in the barrel region of the calorimeters. Their parameters were optimised using a Bayesian procedure that balances energy resolution against network size. Multiple architectures were found to outperform the optimal filtering algorithm currently in use, while remaining small enough to satisfy the stringent hardware and latency constraints of the readout electronics. In particular, the Dense and CNN networks improve the energy resolution by about 8\% while requiring fewer than 500 MAC units, making them suitable for FPGA implementation. The inclusion of pre-deposit samples was found to be essential for capturing the effects of previous energy deposits, which makes recurrent architectures less competitive due to their higher resource requirements for long sequences.

The Dense architecture was further extended with Deep Evidential Regression to provide per-event uncertainty estimates on the predicted energy, opening the way to improved cell energy selection in clustering algorithms. This approach does not significantly increase inference costs and remains well suited for FPGA deployment. The predicted uncertainties were found to be consistent, on average, with the differences between the true and predicted energies. The decomposition of the predicted uncertainty into aleatoric and epistemic components revealed that epistemic uncertainty dominates, suggesting that further gains could be achieved with larger training datasets or refined architectures.

\backmatter

\bmhead{Acknowledgements}
The project leading to this publication has received funding from Excellence Initiative of Aix-Marseille Universit\'e - A*MIDEX, a French ``Investissements d’Avenir'' programme, AMX-18-INT-006 and from the French ``Agence National de la Recherche'', ANR-20-CE31-0013. This work received support from the French government under the France 2030 investment plan, as part of the Excellence Initiative of Aix-Marseille University - A*MIDEX (AMX-19-IET-008 - IPhU). This work was in part supported by the German Federal Ministry of Research, Technology and Space within the research infrastructure project 05H24OD9, and by the German Research
Foundation under project number 460248186 (PUNCH4NFDI).

\section*{Declarations}

\bmhead{Competing interests} The authors have no competing interests to declare.

\bibliography{sn-bibliography}% common bib file

\end{document}